\documentclass[mathleft]{an}
\usepackage{graphicx}
\usepackage{times}
\usepackage{graphicx}
\usepackage{times}
 \def\be{\begin{equation}}
 \def\ee{\end{equation}}
 \def\bea{\begin{eqnarray}}
 \def\eea{\end{eqnarray}}

\begin{document}


\title{Core - Corona Model analysis of the Low Energy Beam Scan at RHIC\fnmsep\thanks{
  Relativistic Heavy Ion Collider in Brookhaven (USA)}}

\author{M.~Gemard
\and  J.~Aichelin \fnmsep\thanks{Corresponding author:
  \email{aichelin@subatech.in2p3.fr}\newline}
}
\institute{
SUBATECH, CNRS/IN2P3, Universit\'e de Nantes, Ecole
  des Mines de Nantes\\
4 rue Alfred Kastler, 44307 Nantes cedex 3, France}



\abstract{%
The centrality dependence of spectra of identified particles in collisions between ultrarelativistic heavy ions with a center of mass energy ($\sqrt{s}$) of 39 and 11.5  $AGeV$ is analyzed in the core - corona model. We show that at these energies the spectra can be well understood assuming that they are composed of two components  whose relative fraction depends on the centrality of the interaction: The core component which describes an equilibrated quark gluon plasma and the corona component which is caused by nucleons close to the surface of the interaction zone which scatter only once and which is identical to that observed in proton-proton collisions. The success of this approach at 39 and 11.5  $AGeV$ shows that the physics does not change between this energy and $\sqrt{s}=200~ AGeV$ for which this model has been developed (Aichelin 2008). This presents circumstantial evidence that a quark gluon plasma is also created at center of mass energies as low as 11.5 $AGeV$. }

\maketitle

\section{Introduction}
There is ample evidence by now that  in heavy-ions collisions at beam energies which can be reached at the colliders at CERN and at RHIC  in Brookhaven a plasma of quarks and gluons is created. Such a state is predicted by 
lattice gauge calculations at high density and/or  temperature.  This plasma is a very short-living state - it lasts less than $10^{-23}$ s-  but it is assumed that this time is sufficiently long for reaching equilibrium. Assuming such an early equilibrium, whose origin is still debated, hydrodynamical calculations describe many details of the observables.

The heavy ion collisions measured at 64 and 200 AGeV at the RHIC accelerator facility in
Brookhaven show a remarkable feature. For the most central collisions the measured multiplicity of all
hadrons which contain up, down and strange quarks only agrees almost perfectly with 
the assumption that the system is at the moment of hadronisation in statistical equilibrium
and that therefore these ratios are determined by a temperature and a chemical potential only.
In addition, the chemical potential is close to zero so therefore the ratios are essentially described
by a temperature. Furthermore this temperature is close to the critical temperature Tc ≈ 152 MeV,
determined as the deflection point of the trace anomaly, for which lattice gauge calculations predict that
the plasma of quarks and gluons goes over into hadronic matter via a cross over (Borsanyi {\it et al.} 2010).
This observation is one of the key evidences that indeed in these collisions a plasma of quarks and gluons is formed.

For symmetric systems the number of projectile participants equals that of target participants, independent of the centrality. If each participant contributes the same energy in the center-of-mass system and if the system comes to equilibrium, one does not expect that the multiplicity per participating nucleon of any type of hadron varies with centrality.  In the experiments such a variation has been observed, however.  In addition the centrality dependence depends strongly on the particle species. Whereas for $\pi$ this ratio is almost constant, for multi strange baryons this ratio varies by a large factor, a phenomenon which has been dubbed strangeness enhancement.

\section{Core-Corona Model}
The basic assumption in statistical model calculations is that geometry does not play a role and {\it all} nucleons
come to statistical equilibrium, means that all phase space configurations compatible with the overall quantum numbers become equally probable.  In simulations of the heavy-ion reactions on an event-by-event basis
(Werner 2007), however, it has been observed that this is not the case. Nucleons close to the surface of the interaction region suffer from less collisions than those in the center of the reaction and there is a non-negligible fraction of nucleons which scatter only once and therefore will not come to an equilibrium with their environment. The relative fraction of these surface nucleons
decreases with centrality. 

This observation has motivated the core-corona model in which it is assumed that nucleons which scatter initially {\it only once} (corona particles)  are not part of the equilibrated source but produce particles as in $pp$ collisions, whereas all the other come to statistical equilibrium (core particles). Of course this fast transition between core and corona particles is a crude approximation but it allows to define from experimental $pp$ and central $AA$ data the centrality dependence of the different observables. Studies have shown that the present quality of data does not allow for a more refined definition of the transition between core and corona particles. The fraction of core, $f_{core}$ and corona particles $f_{corona}=1-f_{core}$ as a function of the centrality has been determined by a Glauber Monte Carlo approach (Aichelin 2008) and is displayed in Fig. 1.
It has been further verified that the core-corona model describes quantitatively the results of the much more involved EPOS simulation program. 
\begin{figure}[h]
\includegraphics[width=0.4\textwidth]{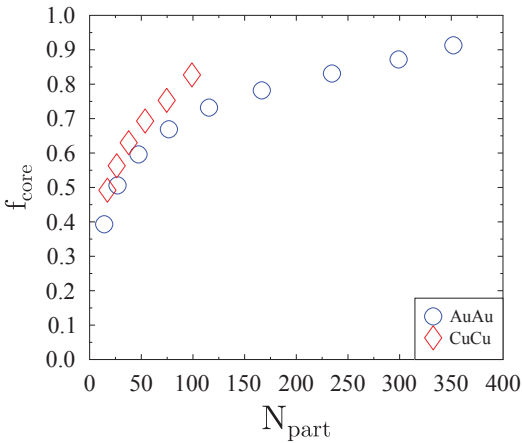} 
\caption{ The core fraction as a function of the centrality expressed by the number of participating nucleons, $N_{part}$ .}
\label{fig0}
\end{figure}
In the core-corona model the centrality dependence of the multiplicity  of a given particle species $i$ in a centrality bin containing $N_{part}$ participants is given by :
\begin{eqnarray}
M^i(N_{part})&=& N_{part}\cdot [f_{core}(N_{part})\cdot M^i_{core} \nonumber \\ 
&+& (1 - f_{core}(N_{part}))\cdot M^i_{corona} ]
\label{eq:M}
\end{eqnarray}
where $M^i_{core}$ is the multiplicity per core participant and $M^i_{corona}$ the multiplicity per corona participant. There are several ways to determinate these two values: one can either calculate them from integrated fits or one can use directly the  published values, which are the results from a fit of a specific form (blast wave model) to the experimental spectra. We chose the latter in the present study: we use the multiplicity measured in pp and divided by a factor of two for $M^i_{corona}$. Then we extract $M^i_{core}$ from the most central multiplicity bin using Eq. (1).

In a series of papers (Aichelin 2008, Aichelin 2010, Aichelin 2010) it has been shown that the core-corona model describes quite nicely the centrality dependence of the multiplicity of identified particles,  $<p_t>$ of identified particles, spectra of protons and $K^+$  (Schreiber 2012) and even of $v_2$ observed in AuAu and PbPb collisions.

\section{Core-Corona Analysis of the Low Energy Beam Scan}
At low beam energies we know that a heavy ion collision can be described by hadronic degrees of freedom and hadronic cross sections only. At beam energies above 64 AGeV the above discussed analysis (Aichelin 2008) presented convincing evidence that a plasma of quarks and gluons is formed. In order to determine the transition energy between both energy domains the Low Energy Beam Scan has been performed at RHIC in which the $p_T$ spectra of particles have been measured down to $\sqrt{s}= 11.5~ AGeV$. As long as the spectra
can be interpreted as a superposition of these both above mentioned components we can assume that a plasma is still created.
If deviations occur the reaction becomes more complex and we have to study the spectra in more detail.
  
If no final state interactions among hadrons take place (and according to the EPOS calculations such collisions change only the
very low momentum part of the spectra) in the core - corona model the  spectra are as well a superposition of two contributions: the core contribution and the corona contribution.  The corona distribution  $ \frac{d^2N_i^{corona}}{2\pi . p_t . dp_t . dy}$ is the measured pp spectra divided by two, the core contribution  $ \frac{d^2N_i^{core}}{2\pi . p_t . dp_t . dy}$ is obtained from the experimental spectra for the most central AA collisions corrected for the corona contribution which is present even in the most central collision and divided by the number of core participants. Then in the core-corona model the spectra for a given centrality is:
\bea
&\frac{d^2M_i}{2\pi . p_t . dp_t . dy}=
N_{part}[ (1 - f_{core}(N_{part}))\frac{d^2N_i^{corona}}{2\pi . p_t . dp_t . dy}\nonumber \\   &+f_{core}(N_{part})\frac{d^2N_i^{core}}{2\pi . p_t . dp_t . dy}].
\eea
Unfortunately in the low energy beam scan proton-proton collisions have not been measured. We therefore have to use the spectrum of the most peripheral reactions together with $f_{core}$ to extract the spectrum for proton-proton collisions.

\begin{figure}[H]
\includegraphics[width=0.4\textwidth]{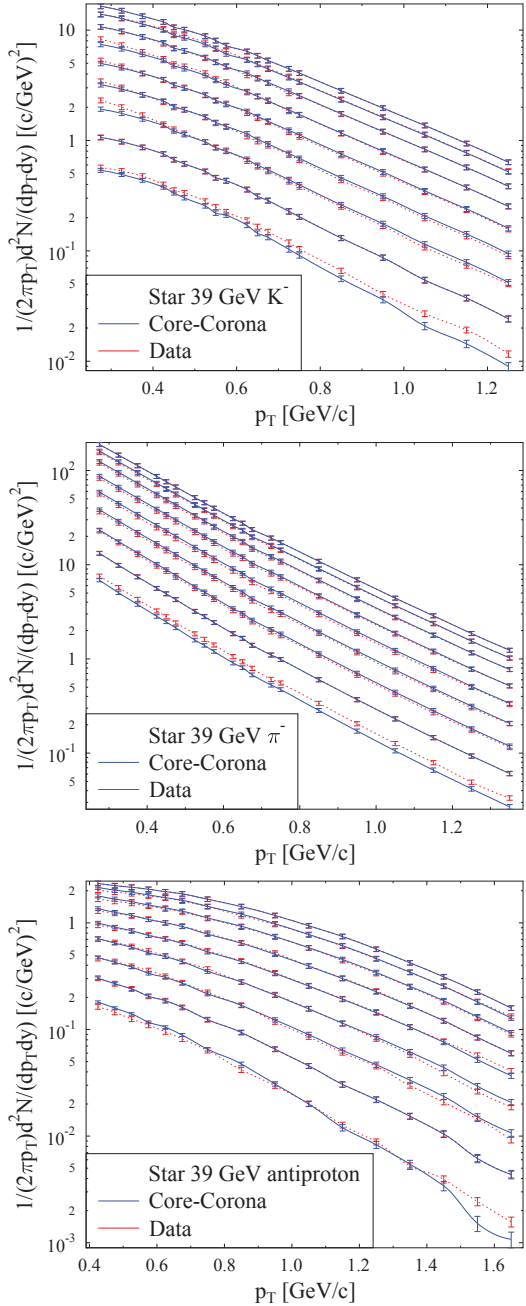} 
\caption{\label{fig1} Centrality dependence of $p_T$ spectra for $K^-$ (top), $\pi^-$ (middle), $\overline{p}$ (bottom) measured by the STAR experiment in $Au+Au$ at $\sqrt{s}=39.~GeV$  (Kumar 2011 and Zhu 2012) in comparison with the predictions of the core-corona model. Here, we calculated $M_{core}$ and $M_{corona}$ with help of the most central and of the second most peripheral collision.}
\end{figure}

The spectrum for the negative charged particles $K^-$ (top), $\pi^-$ (middle) and antiproton (bottom)  (those for the positive charged particles cannot be displayed due to space limitation),  produced in AuAu collisions at $\sqrt{s} = 39~ AGeV $ and $11.5~ AGeV$  (Kumar 2011 and Zhu 2012), are shown for different centralities in Figs. \ref{fig1} and \ref{fig4}, respectively. We observe in both cases a quite nice agreement with the core-corona predictions which are shown in the same graph.

\begin{figure}[H]
\includegraphics[width=0.4\textwidth]{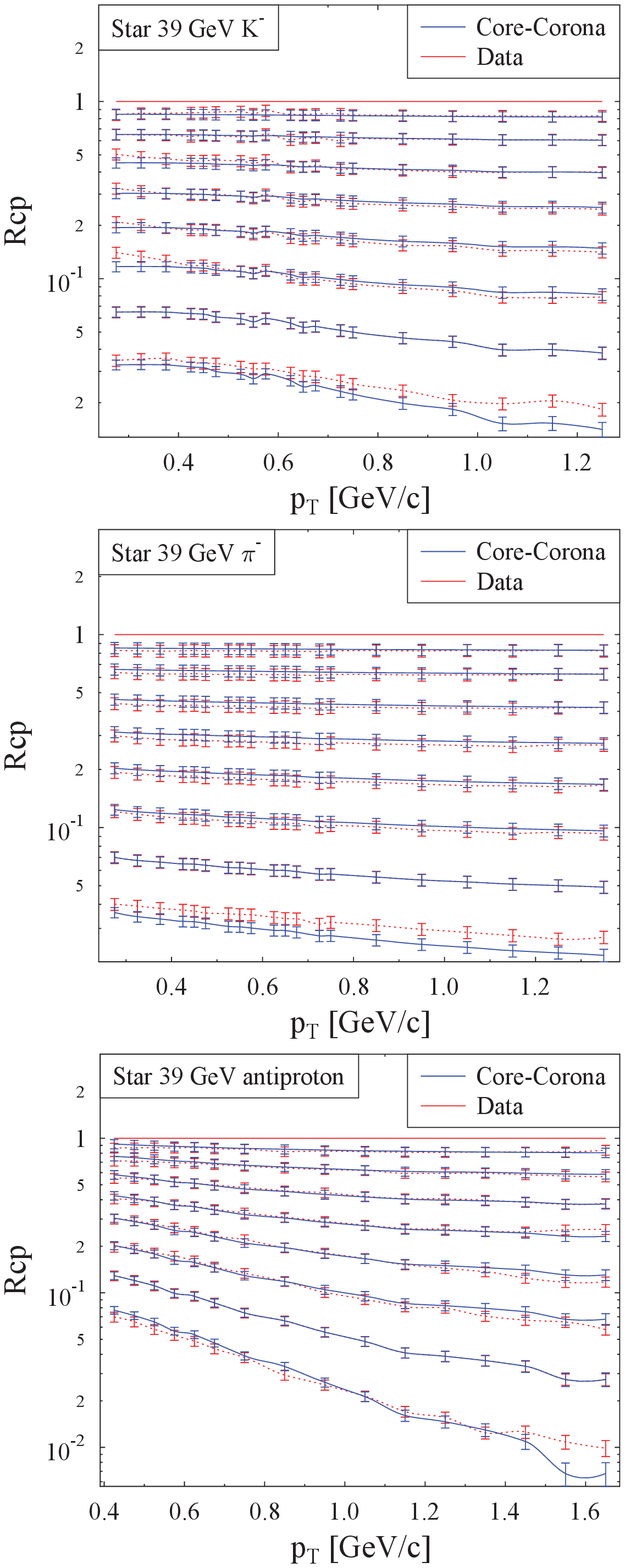} 
\caption{\label{fig2} Centrality dependence of $p_T$ spectra for $K^-$ (top), $\pi^-$ (middle), $\overline{p}$ (bottom) measured by the STAR experiment in $Au+Au$ at $\sqrt{s}=39.~GeV$  (Kumar 2011 and Zhu 2012) in comparison with the predictions of the core-corona model. Here, we display $R_{cp}$, the ratio between the spectrum at a given centrality and the most central spectrum.}
\end{figure}

To quantify this agreement we divide in Figs. \ref{fig2} and \ref{fig5} the spectra at a given centrality by the most central spectrum.
This quantity we dubbed $R{cp}$. This presentation displays clearly whether the slope of the spectra changes as a function of the centrality and whether the core - corona model is able to reproduce this eventual change. We see that the pion and kaon  slope is almost independent of the centrality, only at very peripheral reactions one observes a slight modification. The antiproton
slope, on the contrary, shows a strong centrality dependence. Therefore the antiproton is the most sensitive probe for the validity of the core - corona model because not only the overall magnitude but also the slope has to be reproduced. We see that for all particles the agreement between data and the core - corona model is very good, with the exception of very peripheral reactions at high transverse momentum where the statistics leave also a lot to be desired and the number of participating nucleons has a large error bar. We see as well that the agreement at $\sqrt {s}=39 ~AGeV$ is as good as that
at $11 ~AGeV$.

\begin{figure}[H]
\includegraphics[width=0.4\textwidth]{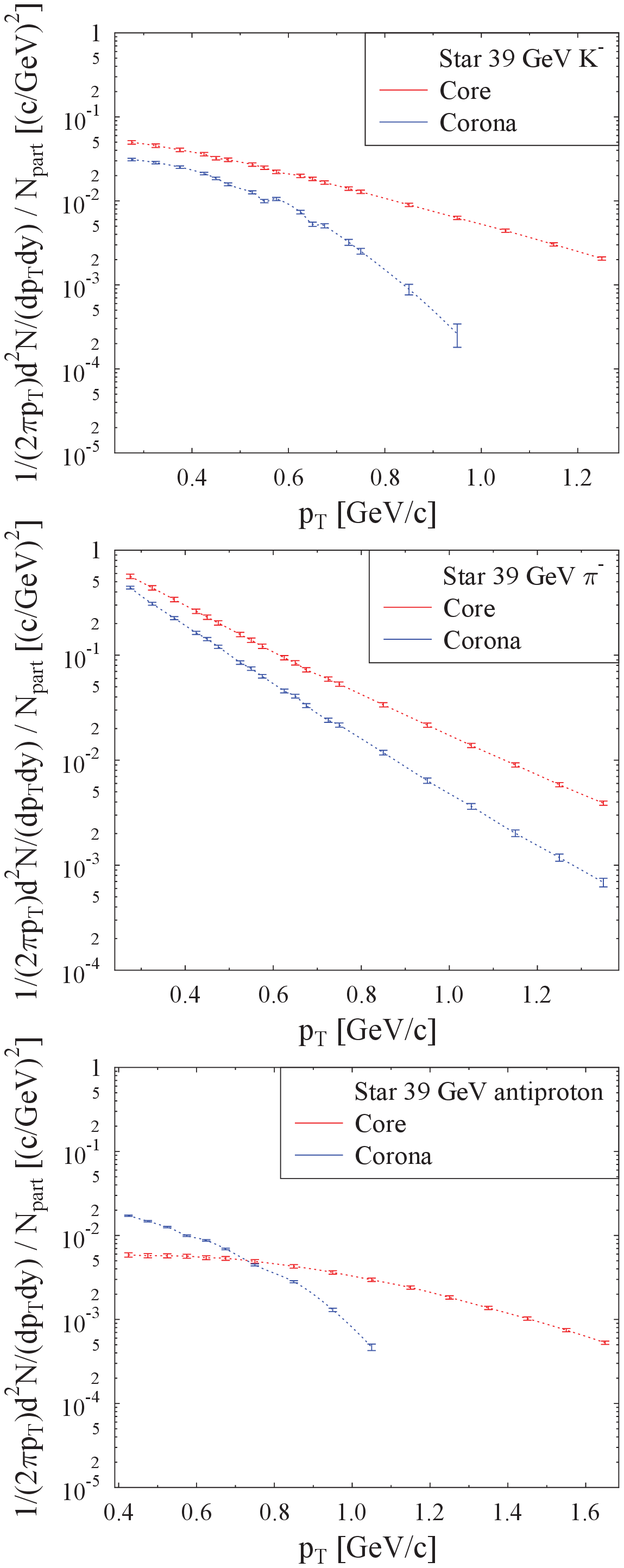} 
\caption{\label{fig3} The $p_T$ spectra for core and corona  and for $K^-$ (top), $\pi^-$ (middle), $\overline{p}$ (bottom) 
for the reaction  in $Au+Au$ at $\sqrt{s}=39.~GeV$ as obtained by the analysis of the spectra in the core - corona model.}
\end{figure}

\begin{figure}[H]
\includegraphics[width=0.4\textwidth]{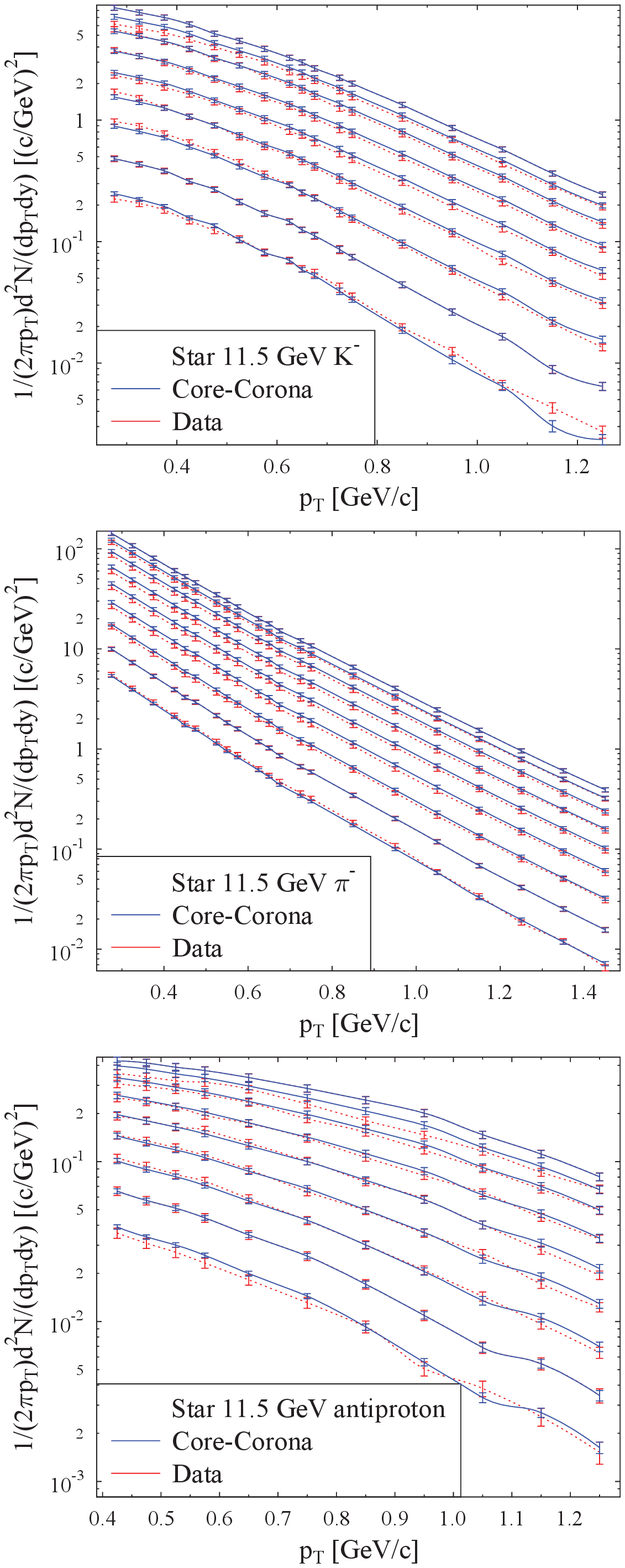} 
\caption{\label{fig4}
 Centrality dependence of $p_T$ spectra for $K^-$ (top), $\pi^-$ (middle), $\overline{p}$ (bottom) measured by the STAR experiment in $Au+Au$ at $\sqrt{s}=11.5~GeV$  (Kumar 2011 and Zhu 2012) in comparison with the predictions of the core-corona model. Here, we calculated $M_{core}$ and $M_{corona}$ with help of the most central and of the second most peripheral collision.}
\end{figure}

\begin{figure}[H]
\includegraphics[width=0.4\textwidth]{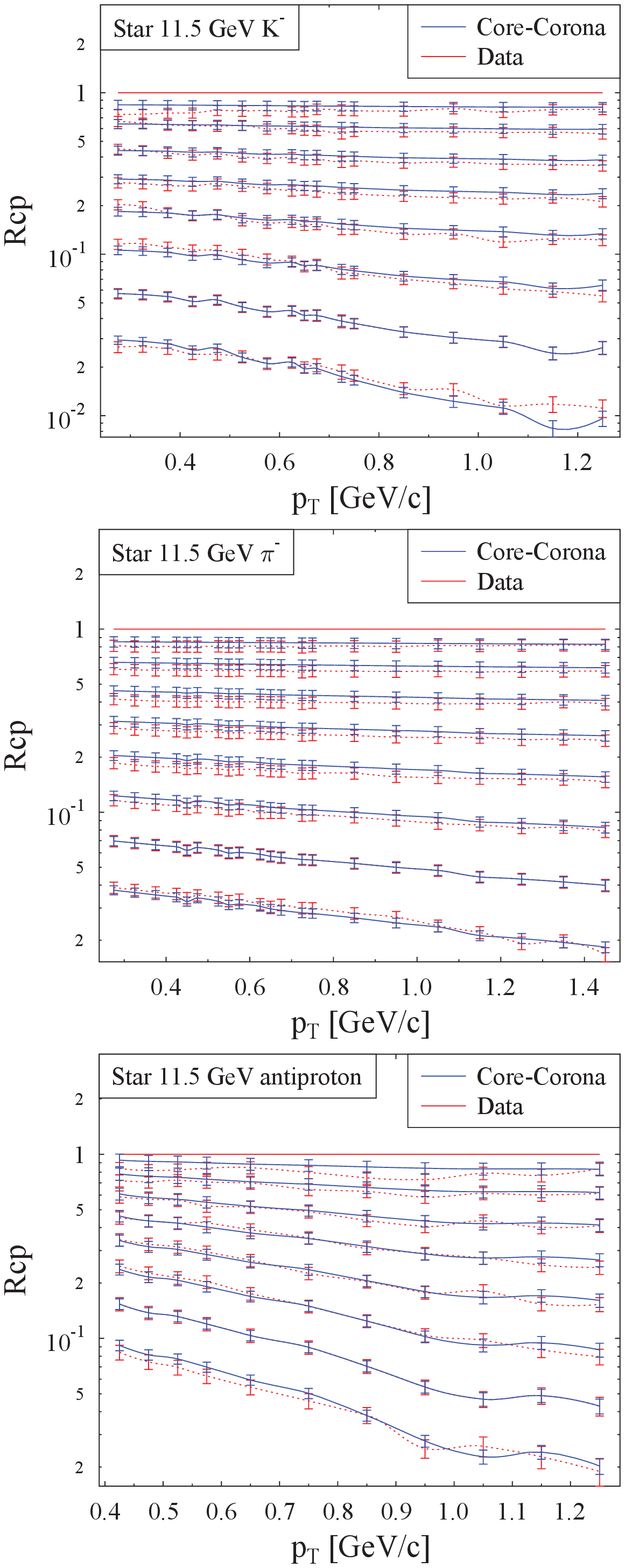} 
\caption{\label{fig5} Centrality dependence of $p_T$ spectra for $K^-$ (top), $\pi^-$ (middle), $\overline{p}$ (bottom) measured by the STAR experiment in $Au+Au$ at $\sqrt{s}=11.5~GeV$ (Kumar 2011 and Zhu 2012) in comparison with the predictions of the core-corona model. Here, we display $R_{cp}$, the ratio between the spectrum at a given centrality and the most central spectrum.}
\end{figure}

\begin{figure}[H]
\includegraphics[width=0.4\textwidth]{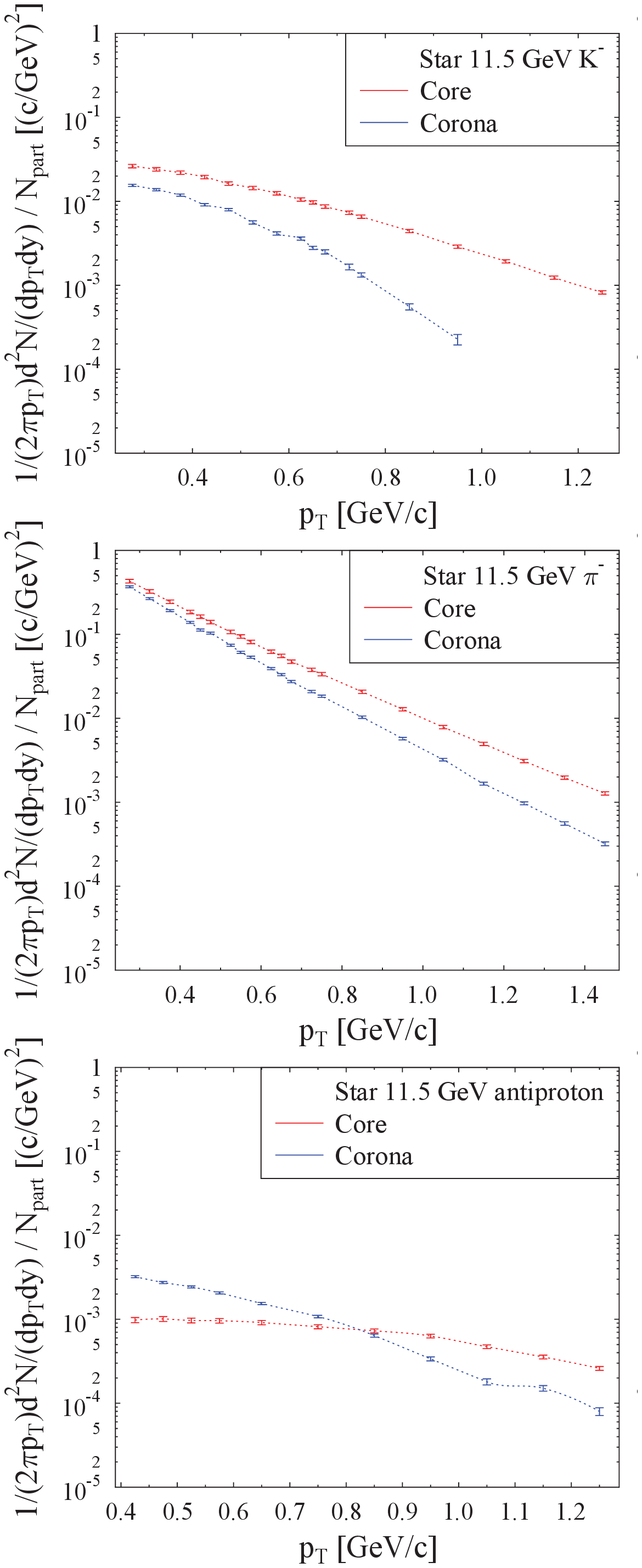} 
\caption{\label{fig6} The $p_T$ spectra for core and corona  and for $K^-$ (top), $\pi^-$ (middle), $\overline{p}$ (bottom) 
for the reaction  in $Au+Au$ at $\sqrt{s}=11.5~GeV$ as obtained by the analysis of the spectra in the core - corona model.}
\end{figure}

Having validated our approach we can now separate $\frac{d^2N_i^{corona}}{2\pi . p_t . dp_t . dy}$ and $\frac{d^2N_i^{core}}{2\pi . p_t . dp_t . dy}$. Both are shown in figs.  \ref{fig3} and \ref{fig6} for the two energies, respectively. They show a quite different form for the core and for the corona particles. The pion spectra are both exponential
and the corona spectrum show a steeper slope. This is understandable because the strings formed in proton-proton collisions 
have on the average not much energy and there high energetic particles are suppressed. For the $K^-$ and the antiprotons
we do not find an exponential form, neither for the core nor for the corona spectrum. Because these particles need associated particles to be produced  ( a $K^+$ for the $K^-$ to conserve strangeness and a proton for the antiproton to conserve the baryon number) the limited available string energy in proton-proton collisions suppresses high transverse momentum $K ^-$ and antiprotons even further than pions. We hope that future proton-proton collisions will verify our approach.

In conclusion, we have shown that the core - corona model is also able to describe heavy ion data at center of mass energies as low as $11.5 ~AGeV$. This presents evidence that the physics does not change substantially between $\sqrt{s} = 200~AGeV $ and $ = 11.5~ AGeV $, especially that there is still a completely equilibrated component in the spectrum which can be identified
with a plasma of quarks and gluons. This has the consequence that the transition region is probably accessible in the presently constructed accelerators at Dubna, Russia  (NICA)  and Darmstadt, Germany (FIAS) and makes the physics in the $\sqrt{s}$ =10 $AGeV$ region even more interesting. 

\acknowledgements
We thank the STAR collaboration for the communication of preliminary data of the low energy s
beam scan and Drs. B. Mohanty,
A. Schmah and X. Dong for fruitful discussions.

\vspace*{2cm}


\end{document}